\documentclass[lettersize,journal]{IEEEtran}
\usepackage{amsmath,amsfonts}
\usepackage{algorithmic}
\usepackage{algorithm,bm}
\usepackage{array}
\usepackage{caption}
\usepackage{textcomp,subcaption}
\usepackage{stfloats}
\usepackage{url}
\usepackage{verbatim}
\usepackage{graphicx}
\usepackage{cite,color}
\usepackage{xcolor}
\begin{document}

\title{Topocode: Topologically Informed Error Detection and Correction in {Image} Communication Systems}

\author{Hongzhi Guo,~\IEEEmembership{Senior Member,~IEEE}
\thanks{H. Guo is with the School of Computing, University of Nebraska-Lincoln, NE, 68588. Email: hguo10@unl.edu}
}

\markboth{Journal of \LaTeX\ Class Files,~Vol.~14, No.~8, August~2021}%
{Shell \MakeLowercase{\textit{et al.}}: A Sample Article Using IEEEtran.cls for IEEE Journals}


\maketitle

\begin{abstract}
Immersive communication technologies such as eXtended Reality (XR) and holographic-type communications require high-data-rate low-latency services. Due to the massive amount of transmitted data, bit-level metrics cannot comprehensively evaluate the Quality-of-Experience (QoE) in these scenarios. This letter proposes Topocode which leverages Topological Data Analysis (TDA) and persistent homology to encode topological information for message-level error detection and correction in image communication systems. It introduces minimal redundancy while enabling effective data reconstruction, especially in low Signal-to-Noise Ratio (SNR) scenarios. Topocode offers a promising approach to meet the demands of next-generation communication systems prioritizing message-level integrity.     
\end{abstract}


\section{Introduction}
Channel coding for error detection and correction is essential in data communication to ensure reliability and achieve high effective data rates. Various codes, such as linear block codes, cyclic redundancy check codes, and convolutional codes, have been extensively developed and applied \cite{proakis2008digital}. These codes primarily address binary source data and enable high-fidelity decoding even in the presence of channel distortion and noise. While accurate binary decoding is desirable, it is important to note that not all bits are equally significant, and certain errors may have minimal impact post-decoding. For instance, in eXtended Reality (XR) and Holographic-Type Communication (HTC)\cite{akyildiz2022holographic,akyildiz2023mulsemedia}, point cloud data with large sizes is common. Communication errors may result in slight shifts in point locations; however, if the overall geometry remains unaffected, the errors are often imperceptible. 

Semantic communication encodes source data by leveraging domain knowledge and its semantic meaning \cite{xie2021deep}. Joint source and channel coding in semantic communication utilizes Deep Learning (DL) models to extract and encode semantic information effectively. At the receiver side, this highly compressed semantic information is decoded using DL decoding models. While semantic communication efficiently encodes source data, {it utilizes application-specific DL models}. Training a unified model for arbitrary data transmission remains a significant challenge. Moreover, DL requires extensive training data, which is not always available, and the trained models often lack generalization. Thus, there is a need for a novel approach that avoids training, applies to any data modality, and focuses on detecting and correcting critical errors without addressing all errors indiscriminately.

In this letter, we propose Topocode, a novel approach leveraging Topological Data Analysis (TDA) \cite{schrader2023topological,hensel2021survey}, particularly persistent homology and Persistence Diagram (PD), to extract and represent {image data's topological information}. {This approach has not been explored in existing coding schemes.} TDA focuses on the shape and geometry of the data, capturing high-level relations and features. Similar to semantic meaning, meaningful data often exhibits underlying manifold structures that can be extracted through TDA. In Topocode, the encoded PD is concatenated with the source data and transmitted through the communication channel. At the receiver, the decoded data is used to generate a new PD, which is compared with the received one. This allows for error detection in the topological space, evaluation of error significance, and correction of topological errors through optimization. {The advantages of Topocode are as follows. First, Topocode introduces minimal redundancy to the source data while effectively detecting message-level errors and assessing their significance. Second, it can correct topological errors, particularly in low Signal-to-Noise Ratio (SNR) scenarios. Third, Topocode utilizes established Computational Topology tools which eliminates the need for training or pre-collected data. This enables its application across various data modalities besides images, such as point cloud data, time-series data, and text, which has broad generalizability.} 

\section{Topological Data Analysis}
\begin{figure}[t]
    \centering
    \begin{subfigure}[b]{0.15\textwidth}
        \centering
        \includegraphics[width=\textwidth]{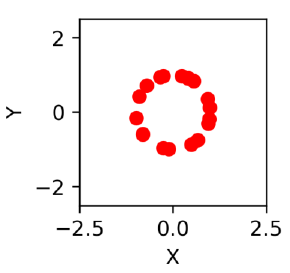}
        \caption{Circle.}
        \label{fig:1-1}
    \end{subfigure}
    \begin{subfigure}[b]{0.15\textwidth}
        \centering
        \includegraphics[width=\textwidth]{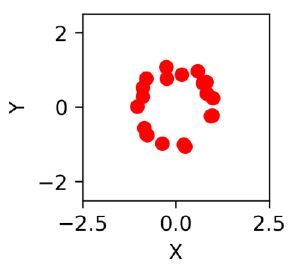}
        \caption{Noisy circle.}
        \label{fig:1-2}
    \end{subfigure}
    \begin{subfigure}[b]{0.15\textwidth}
        \centering
        \includegraphics[width=\textwidth]{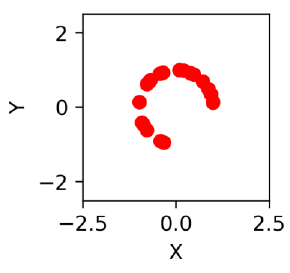}
        \caption{3/4 circle.}
        \label{fig:1-3}
    \end{subfigure}
    \begin{subfigure}[b]{0.15\textwidth}
        \centering
        \includegraphics[width=\textwidth]{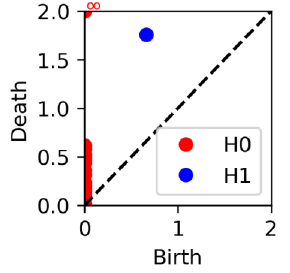}
        \caption{PD of the circle.}
        \label{fig:1-4}
    \end{subfigure}
    \begin{subfigure}[b]{0.15\textwidth}
        \centering
        \includegraphics[width=\textwidth]{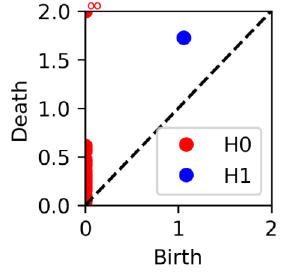}
        \caption{PD of noisy circle.}
        \label{fig:1-5}
    \end{subfigure}
    \begin{subfigure}[b]{0.15\textwidth}
        \centering
        \includegraphics[width=\textwidth]{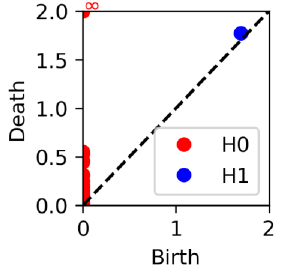}
        \caption{PD of 3/4 circle.}
        \label{fig:1-6}
    \end{subfigure}
    \caption{Comparison of point clouds and PDs for circle variants. }
    \vspace{-15pt}
    \label{fig:point_cloud}
\end{figure}

TDA is used to study the topological features across various data modalities. It identifies different homology groups, such as the \(0^{\text{th}}\) group (\(H_0\)), representing connected components, the \(1^{\text{st}}\) group (\(H_1\)), representing holes or loops, and the \(2^{\text{nd}}\) group (\(H_2\)), representing voids. TDA measures the size of each homology group, providing insights into the overall shape and structure of the data. However, size alone cannot capture the evolution of these homology groups or their relative relationships. To address this limitation, persistent homology is employed to track the evolution of each component in a homology group through Birth-Death (BD) pairs which can capture richer topological information. For example, a point cloud circle and its corresponding PD are shown on the left side of Fig.~\ref{fig:point_cloud}. In the \(H_0\) homology group, BD pairs are born at 0 and die at various values because each point starts as an individual connected component. The birth and death values represent the radii of disks centered at the points. As these disks grow, when two disks touch, one component dies, while the other persists. In the \(H_1\) homology group, a BD pair \((0.66, 1.76)\) represents the circular structure. The circle is formed when the radii of the disks reach approximately 0.66, and as the radii increase to 1.76, the circular structure closes. This process is called the Vietoris-Rips Filtration, where the radius serves as the filtration value \cite{schrader2023topological,hensel2021survey}.

In the middle and on the right-hand side of Fig.~\ref{fig:point_cloud}, we show two circular point clouds which are distorted mildly and severely, respectively. This can be reflected in the PD. First, in Fig.~\ref{fig:1-5}, the \(H_1\) homology group BD pair shifts more towards the diagonal which locates at (1.06,1.72). Since death must be after birth, all BD pairs are above the diagonal. However, if a BD pair is close to the diagonal, it means the BD pair quickly dies after its birth which represents insignificant topological feature. Thus, the circular structure is weaker compared with the one in Fig.~\ref{fig:1-1}. Second, for the point cloud in Fig.~\ref{fig:1-3}, the \(H_1\) homology group is almost on the diagonal which locates at (1.69,1.77) in Fig.~\ref{fig:1-6}. This shows that the circular structure is not strong which can be neglected, which indicates a significant difference from the one in Fig.~\ref{fig:1-1}. 

Since persistent homology captures the intrinsic topological information in the data, the size of the homology group depends on the data which is not a constant. When the number of BD pairs becomes large, it is challenging to evaluate the change of PD. In persistent homology, there are well-defined non-Euclidean distances. Consider that the source data is ${\bm X}$ and the distorted data is ${\hat {\bm X}}$. Then, we perform filtration using a function $f_{l}(\cdot)$ (such as Vietoris-Rips Complex) and obtain PD $D(f_{l}(\cdot))$. The difference between ${\bm X}$ and ${\hat {\bm X}}$ can be evaluated using their PDs \cite{kerber2017geometry},
\begin{align}
    d_{w,p} (D(f_{l}({\bm X})),D(f_{l}(\hat {\bm X})))=\nonumber\\
    \left(\sum_{x \in D(f_{l}({\bm X})) \cup \Delta}\|x-\eta(x)\|_{\infty}^p\right)^{1 / p}
\end{align}
where $\Delta$ is the diagonal of the PD, $\|(a,b)\|_{\infty}=\max \{|a|,|b|\}$ and $\eta(x)$ is a bijection function which maps the BD pairs in $D(f_{l}({\bm X}))$ to those in $D(f_{l}(\hat {\bm X}))$, i.e., $\eta: D(f_{l}({\bm X})) \cup \Delta \rightarrow D(f_{l}(\hat {\bm X})) \cup \Delta$. Since the two diagrams may have different number of BD pairs, some of the BD paris that cannot be matched are mapped to the diagonal $\Delta$. The above distance measurement is called $p$-Wasserstein distance.

\begin{figure}[t]
    \centering
    \includegraphics[width=0.24\textwidth]{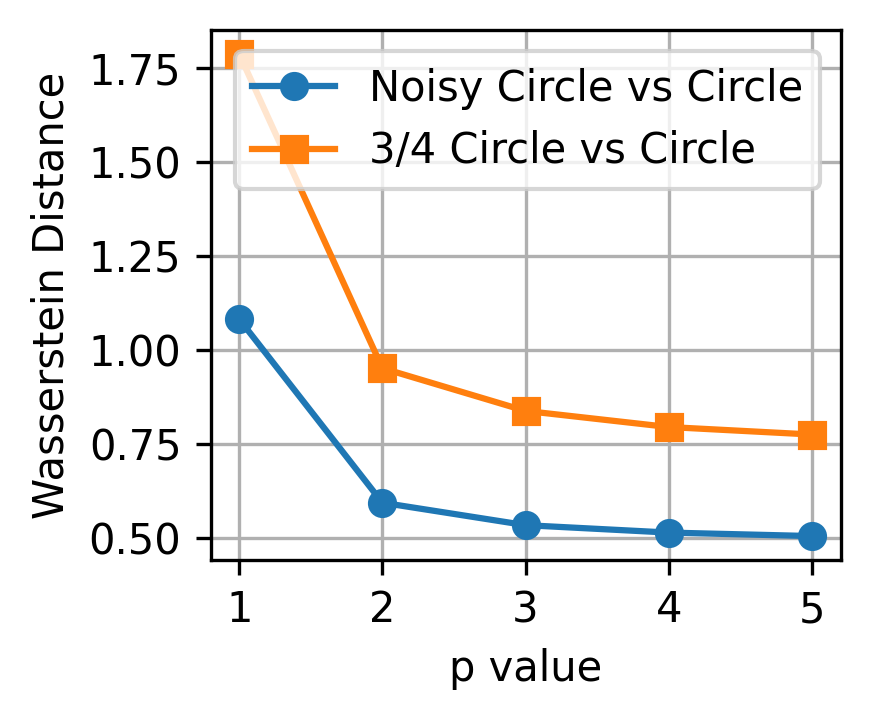}
    \vspace{-10pt}
    \caption{Wasserstein distance between the examples in Fig.~\ref{fig:point_cloud} with different $p$ values. }
    \vspace{-15pt}
    \label{fig:wdistance}
\end{figure}%

\begin{figure}[t]
    \centering
    \includegraphics[width=0.23\textwidth]{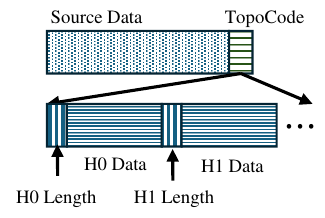}
    \vspace{-7pt}
    \caption{Illustration of the proposed Topocode data packet structure.}
    \vspace{-18pt}
    \label{fig:topocode}
\end{figure}
An example of the $p$-Wasserstein distance between the point clouds in Fig.~\ref{fig:point_cloud} is shown in Fig.~\ref{fig:wdistance}. The $p$-Wasserstein distance includes the summation of the \(H_0\) and \(H_1\) distances. The noisy circle and the 3/4 circle are compared with the original circle. As we can see, the noisy circle has a smaller $p$-Wasserstein distance when $p$ is from 1 to 5. In this paper, without specific notation, we use $p=2$, which is a widely used parameter. Note that, the Wasserstein distance between two identical diagrams is 0. Thus, the distance can be used to evaluate the received data's topological fidelity.

\section{Error Detection and Correction}
In this section, we introduce the Topocode structure and the algorithms for error detection and correction using PDs. {An illustration of the encoded Topocode structure is shown in Fig.~\ref{fig:topocode}. The source data, which is an image or a segment of image, is uncoded.} The packet meta data can contain information about the source data length. Topocode is concatenated with the source data and it is organized by the order of homology groups. Each homology group contains its length and the homology group data. Although only \(H_0\) and \(H_1\) are shown, the homology group number can be {varied depending on the application, e.g., 3D point cloud data may use \(H_2\)}.

The homology group data, e.g., \(H_0\) and \(H_1\) data, can have many different formats. First, the PD can be directly encoded using the coordinates of BD pairs. The encoded homology group data is a series of BD pairs. {However, the number of BD pairs is unknown which results in variable code length. To address this issue, there are various other fixed-length topological features that can be derived from PDs, such as the persistence landscape, persistence image, quantized PD, and amplitude-based features \cite{hensel2021survey}. Although they can generate fixed-length code, we lose complete information of the original PD and topological distance metrics such as Wasserstein distance cannot be used. Since this is the first paper uses topological information for error detection and correction, we use the original PD which contains all available topological information}. Other fixed-length or quantized topological features will be studied in future works.      

{Different data modalities can be processed by specific filtration approaches due to unique data formats. For example, Cubical Complex is more efficient for images due to their grid structures, and Alpha Complex and Vietoris-Rips Complex are suitable for Point Clouds since the points are distributed in 3D space without any specific format. This letter only considers gray images and uses Cubical Complex to obtain the PDs \cite{gudhiurm}.} 

In persistent homology, the stability of PD for grayscale functions is ensured in the following way \cite{skraba2020wasserstein}:
\begin{align}
\label{equ:distance}
    d_{w,p} (D(f_{l}({\bm X})),D(f_{l}(\hat {\bm X})))\leq C_{fil} \|{\bm X}-{\hat{\bm X}}\|_p 
\end{align}
where $C_{fil}$ is a coefficient which depends on the filtration types and the data values. The above inequality shows that $C_{fil}$ times the difference of two gray images is no smaller than their $p$-Wasserstein distance. If two images are similar, the $p$-Wasserstein distance must be small. As a result, the topological errors in the received data can be detected using PDs. 
\begin{figure}[t]
    \centering
    \includegraphics[width=0.22\textwidth]{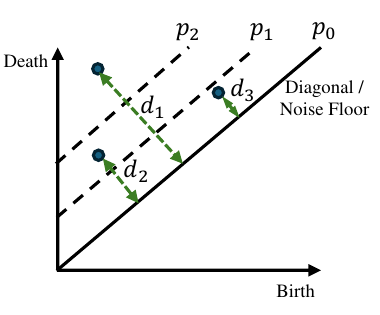} 
    \vspace{-10pt}
    \caption{Illustration of topological total persistence. }
    \vspace{-15pt}
    \label{fig:error_detection}
\end{figure}

Equation~\eqref{equ:distance} provides a lower bound of the distance between ${\bm X}$ and ${\hat {\bm X}}$. If we change ${\hat {\bm X}}$ to noises, Eq.~\eqref{equ:distance} shows the topological information in ${\bm X}$, i.e., the distance to noises in the topological space. As we have discussed, the noises in PDs mainly appear around the diagonal, which can be considered as the noise floor. As shown in Fig.~\ref{fig:error_detection}, a PD contains three BD pairs with distances to diagonal being $d_1$, $d_2$, and $d_3$. If we increase the noise floor from $p_0$ to $p_1$, the BD pair 3 is considered as noises which is not important and can be removed. Then, we obtain an updated denoised PD with BD pairs 1 and 2. We can further increase the noise floor to $p_2$ and only keep significant BD pairs. In order to select the $p_1$ and $p_2$, we use the total persistence in this letter, which is the summation of all the finite distances of BD pairs to the diagonal
$T_t^h = \sum_{i\in D_h(f_{l}({\bm X}))} d_i, {\text { for }} h \in\{0, 1, \cdots\},$
where $D_h(f_{l}({\bm X}))$ represents the $h$-th homology group in the PD. In this case, we consider the homology groups individually. Each homology group $h$ has its own total persistence $T_t^h$. The total persistence can represent the strength of the topological feature. For instance, Gaussian noises with limited number of BD pairs have nearly 0 total persistence.

For error detection, we compare the received data's PD with Topocode. However, noises can corrupt the transmitted data, and the Wasserstein distance in Eq.~\eqref{equ:distance} can be large. In order to reduce the impact of noises, we remove the insignificant BD pairs, which can be introduced by noises, by choosing a threshold percentage $\alpha\in [0,1]$ and set new noise floor $p^h = \alpha T_t^h$. All BD pairs below $p^h$ are removed and we only compare the significant BD pairs. If removing insignificant BD pairs and the distance between the updated PD is small to the original one, the significant topological feature still remains in the received data. The errors can be accepted without requesting for retransmission. The selection of $\alpha$ depends on how much error can be tolerated. We can gradually increase $\alpha$ to evaluate the similarity at different persistence levels. The $1-\alpha$ can be considered as the overall significance level. The larger $\alpha$, the smaller overall significance. 

Finally, we can obtain the Wasserstein distance at different significance level which represents the detected error significance. For example, if the Wasserstein distance decreases as $\alpha$ increase, that means most of the errors are insignificant and the major topological feature of the data remains. On the contrary, if the Wasserstein distance does not change much as $\alpha$ increase, that means the error is significant which may require retransmission.  
Note that, Topocode error detection evaluate the data at the message topology level. {The received data can have insignificant bit-level errors but Topocode cannot detect those.}

Given the received data and Topocode, we can detect their mismatch using the error detection. In addition, we can correct the received data with topological errors using the received Topocode in the following way:
\begin{equation}
\label{equ:correction}
    {\tilde {\bm X}}=\arg \min_{\hat {\bm X}\in {\mathcal D}} \sum _{h}^H d_{w,p,h} (D_h(f_{l}({\bm X})),D_h(f_{l}(\hat {\bm X}))),
\end{equation}
where $H$ is number of homology groups used, ${\mathcal D}$ is the set where the source data ${\bm X}$ is defined. The idea is to minimize the $p$-Wasserstein distance and obtain the recovered source data ${\tilde {\bm X}}$. {In Eq.~\ref{equ:correction}, the $D_h(f_{l}({\bm X}))$ is the Topocode sent by the transmitter.}

In order to solve Eq.~\eqref{equ:correction}, we can use gradient decent. It is essential to ensure that the filtration and the $p$-Wasserstein distance computation are differentiable. The optimization of persistent homology functions has been studied in \cite{carriere2021optimizing}. The PD is formed by BD pairs which are generated by complexes. The complex is generated based on distance functions which are differentiable. As a result, we can solve Eq.~\eqref{equ:correction} using the following steps. {First, ${\hat {\bm X}}$ is normalized by dividing their maximum values before obtaining the PD, so as the ${{\bm X}}$ at the transmitter side before generating Topocode}. Second, given the target PD $D(f_{l}({\bm X}))$, we define the optimization loss function, which is
\begin{align}
    {\mathcal L} = &\gamma \sum_{h=0}^{H} d_{w,p,h} (D_h(f_{l}({\bm X})),D_h(f_{l}(\hat {\bm X})))+ \nonumber \\
    &\sum_{i} \min(|{\hat {\bm X}}_{[i]}|, |1-{\hat {\bm X}}_{[i]}|),
\end{align}
where ${\gamma}$ is a weight to balance the trade-off between $p$-Wasserstein distance loss and the regularization loss, $H$ is the number of homology groups, the second term is used to reduce noises and ${\hat {\bm X}}_{[i]}$ represents the \(i^{th}\) item in ${\hat {\bm X}}$. Note that, the selections of $\gamma$ and the type of regularization function have impact on the error correction result. We can minimize ${\mathcal L}$ and obtain ${\tilde {\bm X}}$ which is the recovered version of ${{\bm X}}$. This process does not require pre-training.

{The complexity to obtain the persistence diagram using Cubical Complex filtration can be achieved by $\Theta\left(3^d n + d 2^d n\right)$, where $n$ is the number of vertices and $d$ is the dimension  \cite{wagner2011efficient}. Then, the Wasserstein distance computation has a worst complexity of ${\mathcal O}(m^2)$, where $m$ is the number of Birth-Death pairs in the persistence diagram. Both Topocode encoding and decoding include the computation of PDs and Wasserstein distance.}

\section{Simulation and Discussion}
In this section, we evaluate the performance of Topocode in error detection and error correction in gray image communication. We use the MNIST and Omniglot datasets. The code is implemented using Gudhi library \cite{gudhiurm}.

\begin{figure}[t]
    \centering
    \begin{subfigure}[b]{0.13\textwidth}
        \centering
        \includegraphics[width=\textwidth]{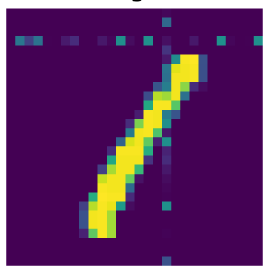}
        \caption{Image-0.}
        \label{fig:5-1}
    \end{subfigure}
    \begin{subfigure}[b]{0.13\textwidth}
        \centering
        \includegraphics[width=\textwidth]{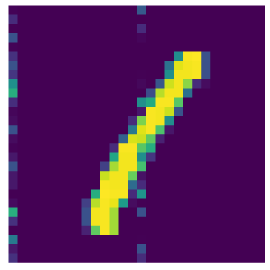}
        \caption{Image-1.}
        \label{fig:5-2}
    \end{subfigure}
    \begin{subfigure}[b]{0.2\textwidth}
        \centering
        \includegraphics[width=\textwidth,height=2.4cm]{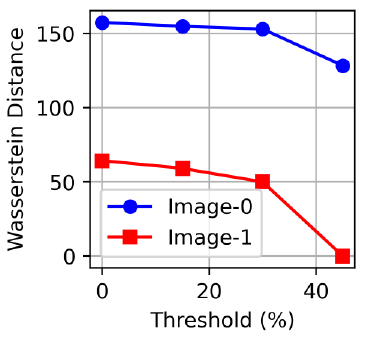}
        \caption{Wasserstein distance.}
        \label{fig:5-3}
    \end{subfigure}
    \caption{Example of error detection using Topocode. The PSNR and SSIM of Image-0 is 24.7 dB and 0.837, respectively. The PSNR and SSIM of Image-1 is 24.7 dB and 0.919, respectively. }
    \vspace{-12pt}
    \label{fig:detection_example}
\end{figure}

First, in Fig.~\ref{fig:detection_example} we show an example using line noises. The Gaussian noise with 1 dB SNR is added to two randomly selected rows or columns. The image data is modulated using BPSK. The received images have the same peak signal-to-noise ratio (PSNR), i.e., 24.7 dB and the structural similarity index measure (SSIM) are 0.837 and 0.919, respectively. As a result, we may consider the noises are similar. However, the noises appear at different locations in the images. In Image-1, a strong noisy column appears on the boundary which does not generate significant impact on the understanding of the image. On the contrary, in Image-0 the noisy columns and rows are close to the digit ``1''. This can be reflected in the Topocode error detection results. In Fig.~\ref{fig:detection_example}, the Wasserstein distance is compared and we vary the threshold of total persistence ($\alpha$) from 0\% to 45\% with an interval of 15\%. As we can see, although the distance of Image-0 drops, it remains a large value. The distance of Image-1 drops to 0 when the threshold is 45\%. This shows that if we remove most of the noises, the recovered image is the same as the source image in topological space. Therefore, although we receive data with errors, Topocode can evaluate the significance of the error.

\begin{figure}[t]
    \centering
    \begin{subfigure}[b]{0.24\textwidth}
        \centering
        \includegraphics[width=\textwidth]{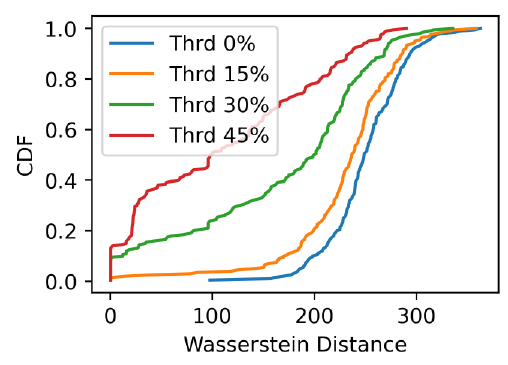}
        \caption{Gaussian noises.}
        \label{fig:6-1}
    \end{subfigure}
    \begin{subfigure}[b]{0.24\textwidth}
        \centering
        \includegraphics[width=\textwidth]{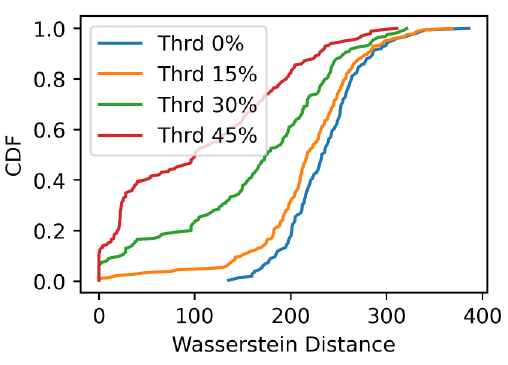}
        \caption{Line noises.}
        \label{fig:6-2}
    \end{subfigure}
    \caption{Impact of $\alpha$ on error detection using Topocode with SNR=3dB.}
    \vspace{-15pt}
    \label{fig:detection_simulation}
\end{figure}

In Fig.~\ref{fig:detection_simulation}, we simulate the transmission of 200 MNIST images with two types of noises. In Gaussian noises, 25\% of pixels are randomly selected and noises with SNR of 3 dB is added to these pixels. In line noises, 25\% of rows or columns are selected and Gaussian noises with SNR of 3 dB is added to these pixels. The PSNR and SSIM are similar for the two types of noise which cannot provide more information about importance of the errors. For received images with Topocode, we set $\alpha$ as 0\%, 15\%, 30\%, and 45\%. Although the noisy images are different from the source images in topological space, after we filter out the insignificant noises, about 20\% of the images have nearly the same topology as the source images. There are about 10\% images whose topology cannot be recovered even we increase the threshold to 45\%, we don't see significant drop of the Wasserstein distance which means the data topology has been significantly altered by noises. 

Next, we show an example of error correction using MNIST dataset. We compare Topocode with existing error correction codes including Low-Density Parity-Check (LDPC) and Convolutional codes. {The regular LDPC code uses a codeword length of 1000 and degrees of (5,20). The code rate of Convolutional code is 2/3.} For LDPC and Convolutional codes, the image is converted to binary data, encoded, modulated using BPSK and added with Gaussian noises and the SNR is 3dB. Then, the received data is demodulated, decoded and converted back to decimal numbers. For Topocode, the PD is obtained using the image and concatenated with the binarized image. Then, the image is modulated using BPSK and added with Gaussian noises with the same SNR. The received data is demodulated and converted back to decimal numbers. Then, the error correction solution in the previous section is used to recover the image.

\begin{figure}[t]
    \centering
    \begin{subfigure}[b]{0.14\textwidth}
        \centering
        \includegraphics[width=2.1cm,height=2.1cm]{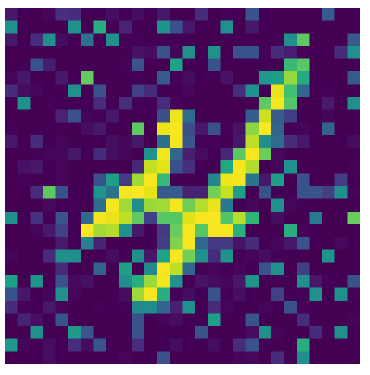}
        \caption{Uncoded image.}
        \label{fig:7-1}
    \end{subfigure}
    \begin{subfigure}[b]{0.11\textwidth}
        \centering
        \includegraphics[width=2.1cm,height=2.1cm]{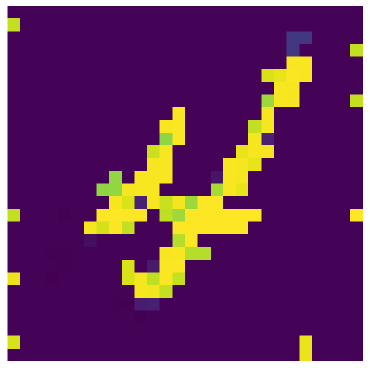}
        \caption{Topocode.}
        \label{fig:7-2}
    \end{subfigure}
    \begin{subfigure}[b]{0.2\textwidth}
        \centering
        \includegraphics[width=2.1cm,height=2.1cm]{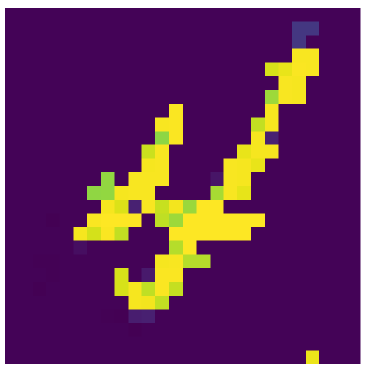}
        \caption{Topocode w/o boundary.}
        \label{fig:7-3}
    \end{subfigure}
        \begin{subfigure}[b]{0.12\textwidth}
        \centering
        \includegraphics[width=2.1cm,height=2.1cm]{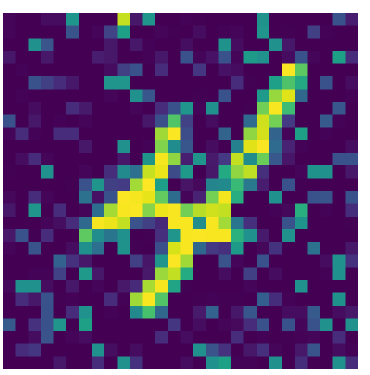}
        \caption{LDPC.}
        \label{fig:7-4}
    \end{subfigure}\quad
    \begin{subfigure}[b]{0.18\textwidth}
        \centering
        \includegraphics[width=2.1cm,height=2.1cm]{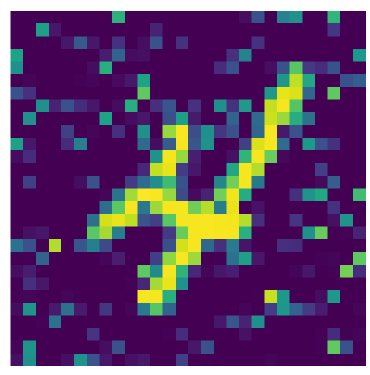}
        \caption{Convolutional code.}
        \label{fig:7-5}
    \end{subfigure}
    \caption{Example of recovered image using Topocode, LDPC, and Convolutional Code with SNR=3 dB. }
    \vspace{-18pt}
    \label{fig:code_compare}
\end{figure}

\begin{figure*}[t]
    \centering
    \begin{subfigure}[b]{0.23\textwidth}
        \centering
        \includegraphics[width=\textwidth]{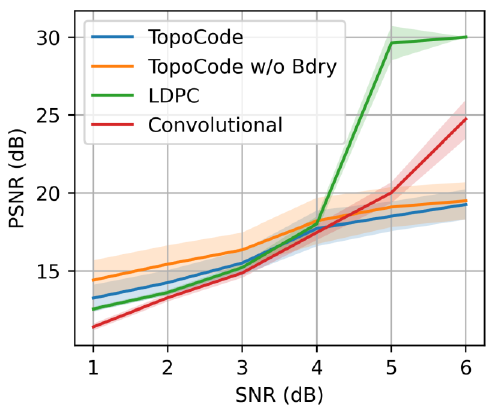}
        \caption{PSNR.}
        \label{fig:8-1}
    \end{subfigure}
    \begin{subfigure}[b]{0.23\textwidth}
        \centering
        \includegraphics[width=\textwidth]{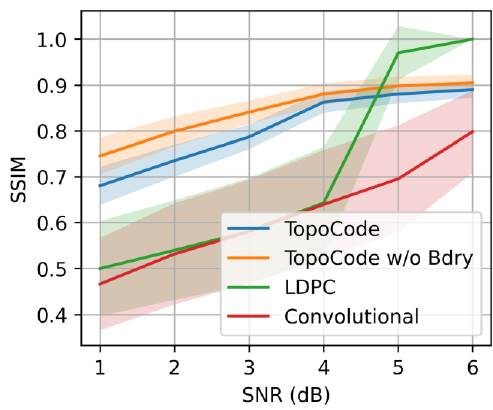}
        \caption{SSIM.}
        \label{fig:8-2}
    \end{subfigure}
    \begin{subfigure}[b]{0.23\textwidth}
        \centering
        \includegraphics[width=\textwidth]{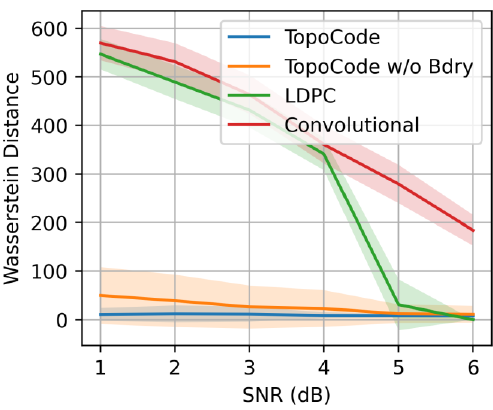}
        \caption{Wasserstein distance.}
        \label{fig:8-3}
    \end{subfigure}
        \begin{subfigure}[b]{0.23\textwidth}
        \centering
        \includegraphics[width=\textwidth,height=3.3cm]{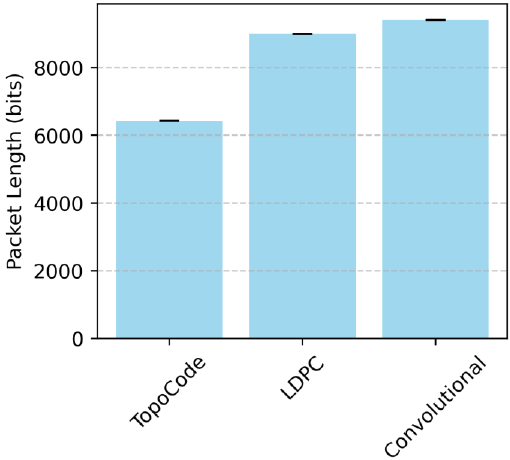}
        \caption{Packet length.}
        \label{fig:8-4}
    \end{subfigure}
    \caption{Error correction performance comparison using MNIST dataset. }
    \vspace{-9pt}
    \label{fig:correction_simulation}
\end{figure*}

\begin{figure*}[t]
    \centering
    \begin{subfigure}[b]{0.23\textwidth}
        \centering
        \includegraphics[width=\textwidth]{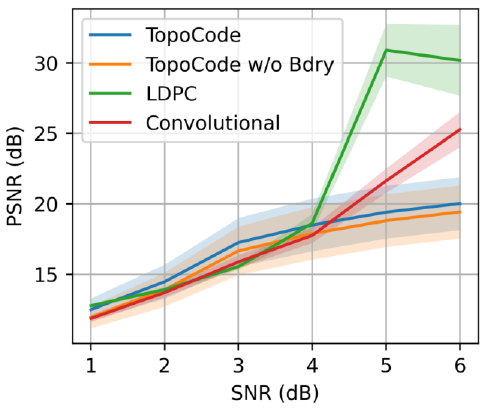}
        \caption{PSNR.}
        \label{fig:8-5}
    \end{subfigure}
    \begin{subfigure}[b]{0.23\textwidth}
        \centering
        \includegraphics[width=\textwidth]{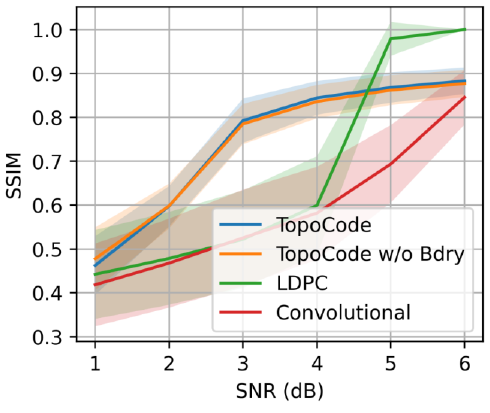}
        \caption{SSIM.}
        \label{fig:8-6}
    \end{subfigure}
    \begin{subfigure}[b]{0.23\textwidth}
        \centering
        \includegraphics[width=\textwidth]{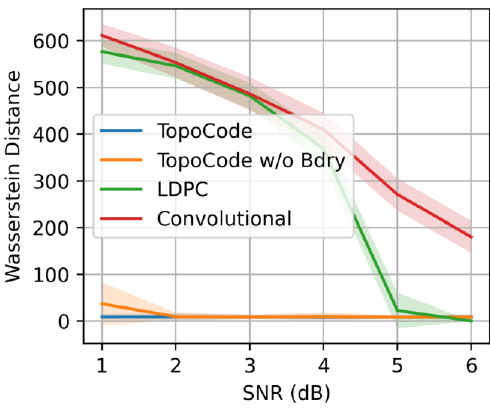}
        \caption{Wasserstein distance.}
        \label{fig:8-7}
    \end{subfigure}
        \begin{subfigure}[b]{0.23\textwidth}
        \centering
        \includegraphics[width=\textwidth,height=3.3cm]{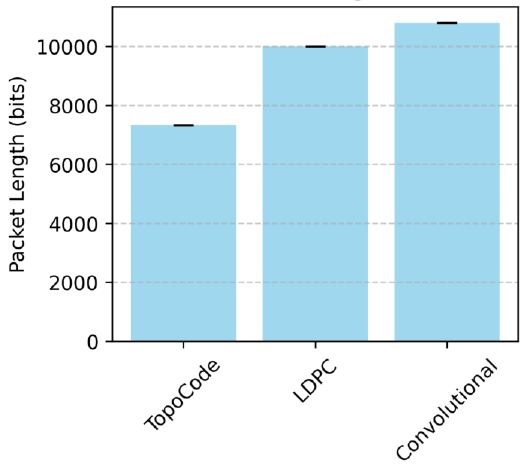}
        \caption{Packet length.}
        \label{fig:8-8}
    \end{subfigure}
    \caption{Error correction performance comparison using Omniglot dataset. }
    \vspace{-15pt}
    \label{fig:correction_simulation2}
\end{figure*}

In Fig.~\ref{fig:code_compare}, we show five recovered images. The uncoded image is the one without any coding and the Topocode is concatenated with uncoded image. The Topocode image is the recovered one after applying the error correction. As we can see, we can remove the majority of noises and recover the topology of the data by optimizing the PD. However, since the Cubical Complex that is used to generate PDs cannot effectively process pixels on the boundary \cite{gudhiurm}, there are remaining noises. In the Topocode without Boundary image, we removed the boundary pixels and the impact of noises further reduces. The decoded images using LDPC and Convolutional Code are also shown which contain significant noises. This example visually shows that Topocode is robust in maintaining data topology in presence of noises, while existing bit-level coding schemes cannot effectively capture data topology or correcting errors based on the content. {Also, there are 15 BD pairs in this example which corresponds to a Topocode length of 256 bits using 8-bit unsigned integer. Compared with the data length $28\times28\times8=6,272$ bits, the code length is only around 4\% of source data.} To further validate the proposed solution, we simulate the transmission of 200 images in MNIST and Omniglot datasets and evaluate the recovered images' PSNR, SSIM, Wasserstein distance, and the code length. The three approaches use the same data and thus the packet length is the length of data and channel coding. In Fig.~\ref{fig:correction_simulation} and Fig.~\ref{fig:correction_simulation2}, we show the results of Topocode, Topocode without boundary, LDPC, and Convolutional code. As we can see from the figure, the Topocode achieves higher PSNR and SSIM performance in the low-SNR regime. As the SNR increases, the LDPC and Convolutional Code perform well. Topocode always shows small Wasserstein distance and short code length. 

{The error correction gain of Topocode at high SNR regime is not as good as that at low SNR regime. In the high SNR regime, only a small number of bit errors exists in the received data. Existing error detection and correction codes can effectively detect and correct them. However, if these errors do not affect the topological information in the data, Topocode cannot detect or correct them. Moreover, another limitation of Topocode is that it can only recover data's topological information, while non-topological information cannot be captured. Although the MNIST and Omniglot datasets contain simple topological information and the PDs are small which leads to short code length, there are various algorithms to reduce the size of Topocode, such as Persistence Landscape, quantification of PD, and Persistence Image \cite{hensel2021survey}. This letter only uses images, but Topocode can be extended to other data modalities, including text, time-series data, and point cloud data. Existing TDA solutions can effectively obtain PDs from these modalities and the error detection and correction are the same as the proposed framework in this letter.} 


\vspace{-5pt}
\section{Conclusion}
In this letter, we propose Topocode which utilizes data topology to detect and correct errors in received data within communication systems. Error detection is achieved through total persistence, enabling the identification of significant errors while filtering out insignificant ones. For error correction, Topocode employs persistence function optimization to restore the data topology using Persistence Diagram (PD). Simulation results demonstrate that Topocode offers unique advantages in error detection and correction at the message level.

\bibliographystyle{IEEEtran} 
\bibliography{topocode} 
\end{document}